\documentclass{article}
\usepackage{spconf,amsmath,amssymb,graphicx}

\usepackage{anyfontsize}

\usepackage{tikz}
\usetikzlibrary{shapes,arrows,fit}
\usetikzlibrary{calc}
\usetikzlibrary{positioning}
\usepackage{tikzscale}
\usepackage{pgfplots}
\pgfplotsset{compat=newest}
\usepackage{graphicx}
\usetikzlibrary{shapes,backgrounds}

\usepackage{bm}

\usepackage{hyperref}
\hypersetup{
    pdfproducer={},  
}

\DeclareMathOperator{\EX}{\mathbb{E}}

\let\oldbibliography\thebibliography
\renewcommand{\thebibliography}[1]{\oldbibliography{#1}
\setlength{\parskip}{0.5pt}%
\setlength{\itemsep}{0.5pt}} 

\title{Speech waveform synthesis from MFCC sequences with generative adversarial networks}

\name{
\begin{tabular}{c}
Lauri Juvela$^1$, Bajibabu Bollepalli$^1$, Xin Wang$^2$, Hirokazu Kameoka$^3$, \\ 
Manu Airaksinen$^1$, Junichi Yamagishi$^2$, Paavo Alku$^1$
\end{tabular}
}

\address{$^1$ Aalto University, Finland  \\
         $^2$ National Institute of Informatics, Japan \\
         $^3$ NTT Communication Science Laboratories, NTT Corporation, Japan 
         }

\begin{document}

{
\onecolumn
\Large 
\noindent \textbf{IEEE Copyright Notice}

\vspace{10pt}

\large
\noindent
\copyright 2018 IEEE. Personal use of this material is permitted. Permission from IEEE must be obtained for all other uses, in any current or future media, including reprinting/republishing this material for advertising or promotional purposes, creating new collective works, for resale or redistribution to servers or lists, or reuse of any copyrighted component of this work in other works. 
\vspace{10pt}

\noindent
Accepted to be Published in: Proceedings of the 2018 IEEE International Conference on Acoustics, Speech, and Signal Processing, April 16-20, 2018, Calgary, Canada
}

\twocolumn

\newpage

\sloppy
\ninept
{
\maketitle
}
\begin{abstract}
This paper proposes a method for generating speech from filterbank mel frequency cepstral coefficients (MFCC), which are widely used in speech applications, such as ASR, but are generally considered unusable for speech synthesis. 
First, we predict fundamental frequency and voicing information from MFCCs with an autoregressive recurrent neural net. Second, the spectral envelope information contained in MFCCs is converted to all-pole filters, and a pitch-synchronous excitation model matched to these filters is trained. Finally, we introduce a generative adversarial network -based noise model to add a realistic high-frequency stochastic component to the modeled excitation signal.
The results show that high quality speech reconstruction can be obtained, given only MFCC information at test time.

\end{abstract}
\section{Introduction}
\label{sec:intro}

Mel freqency cepstral coefficients (MFCCs) \cite{davis1980mfcc} are widely used in speech applications, such as automatic speech recognition (ASR) \cite{rabiner1993fundamentals-asr} and speaker verification (ASV) \cite{kinnunen2010overview,hansen2015speaker-recognition-tutorial}. Since MFCCs are  engineered for these tasks, their use discards lots of signal details that are considered irrelevant in the recognition task. The success of MFCCs in recognition and classification tasks is in part due to this lossy compression, which approximates perceptual properties in hearing \cite{hermansky2013perceptual-properties-asr}. Specifically, MFCCs separate spectral envelope from fine structure, and use a non-linear frequency resolution based on auditory scales. 

Although MFCCs are usually considered sub-optimal for text-to-speech (TTS), and e.g.~mel-generalized cepstrum (MGC) \cite{tokuda1994mel-generalized-cepstrum} is used instead to avoid utilizing filterbanks, reconstructing speech signals from a MFCC representation is sometimes needed. In ASR, for example, understanding causes behind recognition errors or analyzing effects of transcription errors might benefit from conversion of MFCCs to speech. %
Furthermore, state-of-the-art ASR and ASV systems utilizing MFCCs can give rise to novel transformation technologies, such as non-parallel voice conversion based on speaker verification models \cite{kinnunen2017non-parallel-vc-ivector}. Despite being non-ideal for TTS, MFCCs constitute a state-of-the-art representation of speech information in most ASR and ASV systems and therefore conversion of this information to an audible speech waveform is justified.

While high-order MFCCs have been proposed for speech coding \cite{boucheron2012low-bitrate-mfcc-codec}, 
the mel-filterbank sizes and discrete cosine transform (DCT) orders typically used in ASR and ASV result in the speech harmonic structure being smoothed out. In this case, the spectral information contained in MFCCs can be treated as an envelope. 
 Given only this envelope, synthesis of speech requires pitch prediction, i.e.~fundamental frequency (F0) and voicing information must be recovered from the MFCCs. This problem has been studied in a GMM-HMM framework 
\cite{Shao2005-predicting-f0, milner2007prediction-f0}, where F0 and voicing were successfully predicted from a GMM joint distribution with MFCCs. As an obvious extension, modern sequence models, such as recurrent neural networks (RNNs), appear as potential tools for the task. 

The method to convert MFCCs to speech proposed in \cite{milner2007prediction-f0} was relatively simplistic: the recovered spectral amplitude was assumed to be minimum phase and sampled at harmonic frequencies. In the time domain, this corresponds to exciting a minimum phase envelope filter with an impulse train.
This process does not include any aperiodicity in synthesis of voiced speech, and loses the mixed phase characteristics in the excitation of natural speech (i.e.~the glottal flow). 
Recently, neural net -based excitation models have been proposed to generate more realistic excitation waveforms for source-filter vocoding in statistical parametric speech synthesis (SPSS) \cite{juvela2016a-high-pitched-excitation}. Previous work used various acoustic features (such as F0, vocal tract and glottal source envelope parameters 
and harmonic-to-noise ratio), and trained a neural network to map them into a pitch-locked glottal excitation waveform. 
Unfortunately, this type of excitation models are limited due to the point-wise regression in the time domain, which results in smoothing and loss of high frequencies. To overcome this problem, a generative adversarial network (GAN) -based excitation model has been proposed recently \cite{Bollepalli2017-gan-glottal-excitation}. However, GANs commonly suffer from training instability and mode collapse, which we propose to mitigate based on the ideas presented in \cite{Kaneko2017-gan-postfilter} and \cite{Kaneko2017-learned-similarity}.

In this paper, synthesis of speech from MFCCs is studied by presenting three main contributions: first, we show that F0 can be predicted from MFCCs with high accuracy by modifying a recent F0 model \cite{Wang2017-quant-f0}, originally proposed for SPSS. Second, we present an excitation model that maps MFCCs and F0 to excitation waveforms obtained by inverse filtering speech using an MFCC-derived envelope. Finally, we introduce an improved residual GAN-based noise model for generating the high-frequency stochastic component lost in the least-squares excitation model.


The results show that MFCCs can be used to synthesize speech with high quality, when no other information is available at test time. However, we do not advocate the use of MFCCs for envelope modeling in a purpose-built TTS system.
The paper is structured as follows: section \ref{sec:overview} overviews of the synthesis pipeline, 
section \ref{sec:f0_model} describes the mapping of MFCCs to F0, section \ref{sec:mfcc_inversion} details the MFCC to all-pole envelope conversion. 
Section \ref{sec:pulse_model} discusses the excitation pulse model and section \ref{sec:residual_gan} introduces a residual GAN noise model. Section \ref{sec:experiments} describes objective measures and listening experiment, and finally we conclude in section \ref{sec:conclusion}.

\section{Synthesis system}
\label{sec:overview}

\begin{figure}[htb]
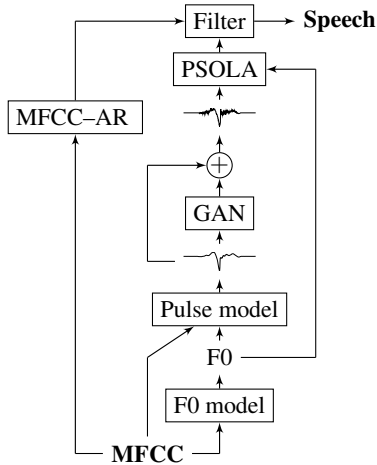

{
\normalsize
\center
\resizebox{0.6\linewidth}{!}{\begin{tikzpicture}
[align=center,node distance=2.0em]
\tikzstyle{rectnode}=[rectangle, minimum width=2em, minimum height=1.0em, draw, fill=white]
\tikzstyle{line}= [-latex', line width=0.15mm]

\node[] (mfcc) {\textbf{MFCC}};
\coordinate [right of = mfcc, node distance = 3em](right_of_mfcc) ; 
\node[rectnode, above of=right_of_mfcc] (f0_model) {F0 model};
\node[ above of=f0_model] (f0) {F0};
\node[rectnode, above of=f0] (pls_model) {Pulse model};
\node[ above of = pls_model, inner sep=3pt] (pls_output) {\resizebox{3em}{!}{\input{figures/genpulse.tikz}}};

\node[rectnode, above of=pls_output] (gan_model) {GAN};
\node[circle, draw, inner sep=0pt, above of= gan_model] (addition) {$+$};
\node[inner sep=3pt, above of=addition] (ref_pulse) {\resizebox{3em}{!}{\input{figures/refpulse.tikz}}};
\node[rectnode, above of=ref_pulse] (psola) {PSOLA};
\node[rectnode, above of=psola] (filter) {Filter};
\node[right of=filter, node distance = 5em] (speech) {\textbf{Speech}};

\node[rectnode, left of=ref_pulse, node distance = 6em] (mfcc-ar) { MFCC--AR };


\draw[] (mfcc) -- (right_of_mfcc);
\draw[line] (right_of_mfcc) -- (f0_model);
\draw[line] (f0_model) -- (f0);
\draw[line] (f0) -- (pls_model.south -| f0);

\coordinate (left_of_f0) at (mfcc |- f0);
\draw[] (mfcc) -- (left_of_f0);
\draw[line] (left_of_f0) -- (pls_model);

\coordinate[ left of=pls_output, node distance = 3em] (left_of_pls) ;
\coordinate (left_of_addition) at (left_of_pls |- addition);
\draw[] (pls_output) -- (left_of_pls);
\draw[] (left_of_pls) -- (left_of_addition);
\draw[line] (left_of_addition) -- (addition);

\coordinate[ right of=psola, node distance = 4em] (right_of_psola) ;
\coordinate (right_of_f0) at (f0 -| right_of_psola) ;
\draw[] (f0) -- (right_of_f0);
\draw[] (right_of_f0) -- (right_of_psola);
\draw[line] (right_of_psola) -- (psola);

\draw[line] (pls_model) -- (pls_output);
\draw[line] (pls_output) -- (gan_model);
\draw[line] (gan_model) -- (addition);
\draw[line] (addition) -- (ref_pulse);
\draw[line] (ref_pulse) -- (psola);
\draw[line] (psola) -- (filter);
\draw[line] (filter) -- (speech);

\coordinate  (left_of_mfcc) at (mfcc -| mfcc-ar) ;
\draw[] (mfcc) -- (left_of_mfcc);
\draw[line] (left_of_mfcc) -- (mfcc-ar);
\coordinate  (left_of_filter) at (filter -| mfcc-ar);
\draw[] (mfcc-ar) -- (left_of_filter);
\draw[line] (left_of_filter) -- (filter);

\end{tikzpicture} }

}
\caption{System overview for MFCC-to-waveform synthesis.}
\label{fig:overview}
\end{figure}

An overview of the synthesis system is shown in Fig.~\ref{fig:overview}. First, F0 is predicted from MFCCs as described in section \ref{sec:f0_model}. Then the MFCCs and predicted F0 is fed into the excitation pulse model detailed in section \ref{sec:pulse_model}. The resulting smooth pulse is further fed into the residual GAN noise model (section \ref{sec:residual_gan}). To create a continuous excitation signal, the generated pulses are joined pitch-synchronously \cite{moulines1995-psola}, as determined by the generated F0. This excitation signal is finally filtered with the envelope reconstructed from the MFCC (section \ref{sec:mfcc_inversion}) to generate a speech waveform.

\subsection{F0 prediction model}
\label{sec:f0_model}

The F0 model takes a sequence of MFCCs as input and generates the corresponding F0 track and voicing information from it. We use a variant of the recently proposed RNN-based model \cite{Wang2017-quant-f0}, which utilizes autoregressive output feedback links and hierarchical softmax for predicting quantized F0 classes from inputs. The F0 range is quantized linearly to 255 bins, and one additional class is reserved for unvoiced speech. In contrast to \cite{Wang2017-quant-f0}, we now use MFCCs instead of linguistic features as inputs, and only have feedback links at frame tier, as no linguistic tier information is available. Network parameters are listed in Table \ref{tab:parameters}.

\subsection{Envelope reconstruction from MFCC}
\label{sec:mfcc_inversion}

This paper utilizes the widely used MFCC computation with HTK-style mel-filterbanks and DCT \cite{young2002htk-book}, as implemented in Librosa \cite{mcfee2015librosa}. 
Spectrum-to-MFCC computation is composed of invertible pointwise operations and linear matrix operations that are pseudo-invertible in the least-squares sense.  This leads to a straightforward reconstruction process: Let the MFCC sequence $\bm C$ be computed as 
\begin{equation}
\bm C = \mathbf{D} \log (\mathbf{M} \bm S) ,
\end{equation}
where $\bm S$  is a pre-emphasized STFT magnitude spectrogram, $\mathbf{M}$ is a mel-filterbank matrix, and $\mathbf{D}$ is a truncated discrete cosine transform matrix.
The reconstruction of the magnitude spectrum is obtained simply by
\begin{equation}
 \hat{\bm S} = \mathbf{M}^+ \exp (\mathbf{D}^+ \bm C) ,
\end{equation}
where $\mathbf{D}^+$ is the pseudoinverse of $\mathbf{D}$ (which coincides with the classical zero-padding and inverse DCT procedure), and $\mathbf{M}^+$ is the pseudoinverse of $\mathbf{M}$. 
Unfortunately, the use of filterbank pseudoinverse does not guarantee non-negativity of the resulting spectrum, but this problem is mitigated by flooring the values to zero \cite{boucheron2008inversion-mfcc}. 
It is possible to instead obtain similar, but always non-negative reconstructions by using interpolation techniques (see e.g.~\cite{kinnunen2017non-parallel-vc-ivector,milner2007prediction-f0}), but we observe that the pseudoinverse behaves well in practice and gives envelopes with a sharper formant structure, compared to the interpolation methods.

An autoregressive all-pole model is fitted to $\hat{\bm S}$ in the conventional manner by computing an autocorrelation from the symmetrized square magnitude via IDFT, and then solving the resulting normal equations (see e.g.~\cite{Makhoul1975} for details). In this paper, we use 24 mel filters, 20:th order MFCCs, and 30:th order all-pole filters, at 16\,kHz sample rate.

\subsection{Excitation pulse model}
\label{sec:pulse_model}

Previously, an excitation model has been proposed for glottal vocoding in SPSS, by using a neural network that maps acoustic features to glottal excitation pulses \cite{juvela2016a-high-pitched-excitation}. A glottal source signal (differential volume flow through the vocal folds) is first obtained with glottal inverse filtering \cite{Airaksinen2014}, after which excitation pulses are extracted by centering the excitation at a pitch mark, cosine windowing a two pitch period segment, and zero-padding the pulse to a fixed length. Finally before training, each acoustic feature frame is associated with a pulse at the nearest pitch mark.  

A similar framework can be adopted generally in all source-filter model -based vocoding, where the filter allows inverse filtering the speech signal. In this paper, we use AR envelopes  reconstructed from MFCCs to approximate the vocal tract filter, and otherwise treat the resulting excitation signals similarly to \cite{juvela2016a-high-pitched-excitation}. Furthermore, the  model input acoustic features are now only MFCCs, log-F0 and voicing information.
Reaper \cite{talkin2015-reaper} is used  to obtain the pitch marks, and only voiced frames are used to train the excitation model.

%

\begin{figure}[htb]
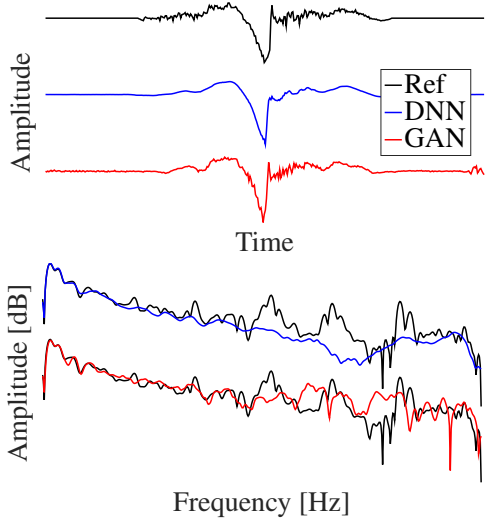


 \centering
  {
   \fontsize{28}{35}
   \resizebox{0.75\linewidth}{!}{\input{figures/pls_plots.tex} }
   \resizebox{0.75\linewidth}{!}{\input{figures/freq_plots.tex} }
  }
  

\caption{Excitation pulses shown in the time and frequency domain. DNN pulse model (blue) fits the overall reference pulse (black) shape, but high-frequency stochastic components are smoothed out, whereas GAN (red) is able to generate realistic high-frequency components.}
\label{fig:signals}
\end{figure}

For the model architecture, we use a gated recurrent unit (GRU) layer at the input, since recurrent nets are powerful for encoding the acoustic sequence information. This is supported by previous research, where recurrent networks slightly improved excitation model performance in a TTS application \cite{juvela2016b-using-text-and-acoustic}. Furthermore, convolution layers have been found convenient when working closer to the waveform level \cite{Bollepalli2017-gan-glottal-excitation}.
As result, we now use a GRU input, followed by a stack of 1D convolution layers, as listed in Table \ref{tab:parameters}.
However, a fundamental limitation in this kind of waveform modeling arises from the point-wise least-squares training criterion. The model will inevitably regress towards a conditional average, given the inputs, which leads to smoothed waveforms and loss of high frequencies. This is illustrated in Fig.~\ref{fig:signals}.

\subsection{Residual GAN model}
\label{sec:residual_gan}

\begin{figure}[htb]
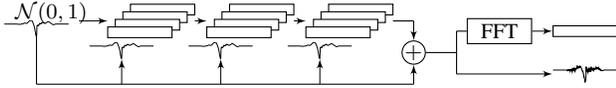

\center
{
\footnotesize
\centering
 \resizebox{0.99\linewidth}{!}{\begin{tikzpicture}
[align=center,node distance=0.5em]
\tikzstyle{rectnode}=[rectangle, inner sep=2pt, minimum width=3em, minimum height=0.1em, draw, fill=white]
\tikzstyle{line}= [-latex', line width=0.1mm]
\coordinate (layer_0) at (0.000000, 0);
\node[rectnode, above right of = layer_0] (node_00) {};
\node[rectnode, below left of = node_00 ] (node_01) {};
\node[rectnode, below left of = node_01 ] (node_02) {};
\node[rectnode, below left of = node_02 ] (node_03) {};
\node [ rotate fit=45, fit=(node_00) (node_03)] (container_0) {};
\node[ inner sep=0pt, below left of = node_03, node distance = 1.2em] (cond_0) {\resizebox{3em}{!}{\input{figures/genpulse.tikz} }};
\coordinate (layer_1) at (1.300000, 0);
\node[rectnode, above right of = layer_1] (node_10) {};
\node[rectnode, below left of = node_10 ] (node_11) {};
\node[rectnode, below left of = node_11 ] (node_12) {};
\node[rectnode, below left of = node_12 ] (node_13) {};
\node [ rotate fit=45, fit=(node_10) (node_13)] (container_1) {};
\node[ inner sep=0pt, below left of = node_13, node distance = 1.2em] (cond_1) {\resizebox{3em}{!}{\input{figures/genpulse.tikz} }};
\coordinate (layer_2) at (2.600000, 0);
\node[rectnode, above right of = layer_2] (node_20) {};
\node[rectnode, below left of = node_20 ] (node_21) {};
\node[rectnode, below left of = node_21 ] (node_22) {};
\node[rectnode, below left of = node_22 ] (node_23) {};
\node [ rotate fit=45, fit=(node_20) (node_23)] (container_2) {};
\node[ inner sep=0pt, below left of = node_23, node distance = 1.2em] (cond_2) {\resizebox{3em}{!}{\input{figures/genpulse.tikz} }};
\coordinate [left of = container_0, node distance = 5em](input_layer);
\node[ above right of = input_layer, node distance = 0.5em] (noise_input) {$\mathcal{N}(0,1)$};
\node[ inner sep=0pt, below left of = input_layer, node distance = 0.5em] (pulse_input) {\resizebox{3em}{!}{\input{figures/genpulse.tikz}}};
\node [ rotate fit=45, fit=(noise_input) (pulse_input)] (input_container) {};
\draw[line] (input_container.east |- input_layer) -- (container_0);
\coordinate[below of = pulse_input, node distance = 2.600000 em] (below_pulse_input);
\draw (pulse_input) -- (below_pulse_input);
\coordinate (below_cond_0) at (below_pulse_input -| cond_0);
\draw [line] (below_cond_0) -- (cond_0);
\coordinate (below_cond_1) at (below_pulse_input -| cond_1);
\draw [line] (below_cond_1) -- (cond_1);
\draw[line] (container_0) -- (container_1);
\draw (below_cond_0) -- (below_cond_1);
\coordinate (below_cond_2) at (below_pulse_input -| cond_2);
\draw [line] (below_cond_2) -- (cond_2);
\draw[line] (container_1) -- (container_2);
\draw (below_cond_1) -- (below_cond_2);
\draw (below_pulse_input) -- (below_cond_0);
\coordinate [right of = container_2, node distance=3em](convstack_output);
\draw (container_2) -- (convstack_output);
\coordinate (below_cond_output) at (below_pulse_input -| convstack_output);
\draw (below_cond_2) -- (below_cond_output);
\node[circle, draw, inner sep=0pt] (addition) at  ($(convstack_output)!0.5!(below_cond_output)$) {$+$};
\draw [line] (below_cond_output) -- (addition);
\draw [line] (convstack_output) -- (addition);
\coordinate [right of = addition, node distance = 2em] (right_of_addition);
\draw (addition) -- (right_of_addition);
\coordinate [above of = right_of_addition, node distance = 1em] (fft_fork);
\draw (right_of_addition) -- (fft_fork);
\node[rectnode, right of = fft_fork, node distance = 2em] (fft_node) {FFT};
\coordinate [below of = right_of_addition, node distance = 1em] (direct_fork);
\draw (right_of_addition) -- (direct_fork);
\node[rectnode, right of = fft_node, node distance = 4em] (fft_output) {};
\node[inner sep=0pt] (direct_output) at (direct_fork -| fft_output) {\resizebox{3em}{!}{\input{figures/refpulse.tikz}}};
\draw [line] (direct_fork) -- (direct_output);
\draw (fft_fork) -- (fft_node);
\draw [line] (fft_node) -- (fft_output);
\end{tikzpicture} }
}
\caption{Generator architecture. Two input channels contain a smooth excitation pulse and white noise, and the pulse is fed to every layer in a residual channel. Additivity at output forces the generator to learn a residual noise-like model. }
\label{fig:generator}
\end{figure}

\begin{figure}[htb]
\center
{
\footnotesize
\centering
  \resizebox{0.99\linewidth}{!}{\begin{tikzpicture}
[align=center,node distance=0.5em]
\tikzstyle{rectnode}=[rectangle, inner sep=2pt, minimum width=3em, minimum height=0.1em, draw, fill=white]
\tikzstyle{line}= [-latex', line width=0.1mm]
\coordinate (layer_0) at (0.000000, 0);
\node[rectnode, above right of = layer_0, minimum width = 2.606420 em] (node_00) {};
\node[rectnode, below left of = node_00, minimum width = 2.606420 em ] (node_01) {};
\node[rectnode, below left of = node_01, minimum width = 2.606420 em ] (node_02) {};
\node[rectnode, below left of = node_02, minimum width = 2.606420 em ] (node_03) {};
\node [ rotate fit=45, fit=(node_00) (node_03)] (container_0) {};
\coordinate [right of = layer_0, node distance = 5.000000 em](layer_1) ;
\node[rectnode, above right of = layer_1, minimum width = 1.371800 em] (node_10) {};
\node[rectnode, below left of = node_10, minimum width = 1.371800 em ] (node_11) {};
\node[rectnode, below left of = node_11, minimum width = 1.371800 em ] (node_12) {};
\node[rectnode, below left of = node_12, minimum width = 1.371800 em ] (node_13) {};
\node [ rotate fit=45, fit=(node_10) (node_13)] (container_1) {};
\coordinate [right of = layer_1, node distance = 5.000000 em](layer_2) ;
\node[rectnode, above right of = layer_2, minimum width = 0.722000 em] (node_20) {};
\node[rectnode, below left of = node_20, minimum width = 0.722000 em ] (node_21) {};
\node[rectnode, below left of = node_21, minimum width = 0.722000 em ] (node_22) {};
\node[rectnode, below left of = node_22, minimum width = 0.722000 em ] (node_23) {};
\node [ rotate fit=45, fit=(node_20) (node_23)] (container_2) {};
\coordinate [right of = layer_2, node distance = 5.000000 em](layer_3) ;
\node[rectnode, above right of = layer_3, minimum width = 0.380000 em] (node_30) {};
\node[rectnode, below left of = node_30, minimum width = 0.380000 em ] (node_31) {};
\node[rectnode, below left of = node_31, minimum width = 0.380000 em ] (node_32) {};
\node[rectnode, below left of = node_32, minimum width = 0.380000 em ] (node_33) {};
\node [ rotate fit=45, fit=(node_30) (node_33)] (container_3) {};
\coordinate [left of = container_0, node distance = 6.000000 em](input_layer);
\node[ rectnode, above right of = input_layer, node distance = 0.5em] (fft_input) {};
\node[ inner sep=0pt, below left of = input_layer, node distance = 0.5em] (pulse_input) {\resizebox{3em}{!}{\input{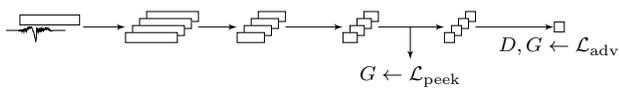}}};
\node [ rotate fit=45, fit=(fft_input) (pulse_input)] (input_container) {};
\draw[line] (input_container) -- (container_0);
\draw[line] (container_0) -- (container_1);
\draw[line] (container_1) -- (container_2);
\draw[line] (container_2) -- (container_3);
\node [rectnode, minimum width = 0.1em, right of = container_3, node distance= 5.000000 em](convstack_output) {};
\node [below of = convstack_output, node distance= 1 em](loss_adv) {$ D,G \leftarrow \mathcal{L}_\mathrm{adv}$};
\draw [line] (container_3) -- (convstack_output);
\coordinate (peek_node) at ($(container_3)!0.5!(container_2)$);
\node [below of = peek_node, node distance = 2.500000 em](peek_output) {$ G \leftarrow \mathcal{L}_\mathrm{peek}$};
\draw [line] (peek_node) -- (peek_output);
\end{tikzpicture} }
}
\caption{Discriminator architecture. Input channels contain time and frequency domain views of the signal. Adversarial loss $\mathcal{L}_\mathrm{adv}$ is used to train both $D$ and $G$. Additionally, $G$ is allowed to peek into discriminator and use the loss $\mathcal{L}_\mathrm{peek}$ to match real and generated data activations. }
\label{fig:discriminator}
\end{figure}

Modeling the aperiodic component of voiced speech in the current synthesis system resembles a previous GAN glottal excitation model \cite{Bollepalli2017-gan-glottal-excitation}, but now the GAN is conditioned on a smooth generated pulse, and the model is forced to generate only an additive residual component.
The training procedure combines LS-GAN \cite{Mao2017lsgan} with GAN-based similarity metric learning \cite{Kaneko2017-learned-similarity}. Furthermore, the residual connections in the generator model are adopted from a GAN postfilter architecture \cite{Kaneko2017-gan-postfilter}. 
Generator and discriminator architectures are shown in Figs.~\ref{fig:generator} and ~\ref{fig:discriminator}, respectively,
with details listed in Table \ref{tab:parameters}.  

Generator input channels are a smooth excitation pulse waveform $\hat{x}$ given by the pulse regression DNN, and white Gaussian noise of the same length. The smooth waveform is further fed into each layer as an additional residual channel. Explicit additivity of $\hat{x}$ and the convolution layers' output ensures that the output resembles residual noise. Finally, a FFT magnitude layer allows the training process to simultaneously see the output in the time and frequency domains, and the error signals can propagate via both routes.

Discriminator input channels see the signal both in the time and frequency domain. Strided convolutions are used to gradually reduce the convolution layer sizes, finally resulting in a single value for classifying the input sample as real of fake. In addition, the generator is allowed to peek into discriminator activations at layer $L$, and use the implied similarity metric to match a generated data mini-batch directly with the corresponding real data.

More formally, the discriminator $D$ attempts to output a "real" value 1 for samples from real data distribution $x \sim p_\mathrm{data}(x)$, and a "fake" value 0 for samples $x^\prime \sim p_G(z \vert \hat{x})$ coming from the generator $G$, where $\hat{x}$ is a smooth output of the previous pulse model, and $z$ is sampled from a standard Gaussian distribution. Similarly to LS-GAN, the discriminator loss function to be minimized is
\begin{equation}
\mathcal{L}_\mathrm{adv}(D) = \frac{1}{2} \EX_{x} \left[ (D(x)-1)^2 \right] 
 + \frac{1}{2} \EX_{x^\prime} \left[ (D(x') )^2 \right] ,
\end{equation}
where $x' =  G(z \vert \hat{x})$.
 Simultaneously, the generator is trained to fool the discriminator to produce the "real" value given generator output
\begin{equation}
 \mathcal{L}_\mathrm{adv}(G) = \frac{1}{2} \EX_{x^\prime} \left[ (D(x')-1)^2 \right] .
\end{equation}
To facilitate learning, the generator is allowed to peek into discriminator activations at layer $L$ and attempt to match its generated output with target data.
\begin{equation}
 \mathcal{L}_\mathrm{peek}(G) = \frac{1}{2} \EX_{x, x^\prime} \left[ (D_L(x) - D_L(x'))^2 \right] .
\end{equation}
This resembles the "learned similarity metric" proposed in \cite{Kaneko2017-learned-similarity}. 

The training procedure alternates between minimizing $\mathcal{L}_\mathrm{adv}(D)$ for discriminator, and $\mathcal{L}_\mathrm{adv}(G) + \mathcal{L}_\mathrm{peek}(G)$ for generator. The discriminator is kept fixed while training the generator. 

\subsubsection{Fourier transform layer}
\label{sec:fft_layer}


In the GAN network, a non-trainable FFT layer is implemented to explicitly output the log spectral magnitude of the input, while allowing backpropagation of error gradients through the layer. 
%
The discrete Fourier transform consists of two differentiable linear operations, given by $F = F_R + i F_I$, where $F_R$ and $F_I$ are the cosine and sine basis matrices. Magnitude is obtained simply by point-wise squaring and summing of the real and imaginary part outputs.
Applying logarithm and scaling pointwise is also differentiable. All of the operations are readily available in Theano, allowing easy integration to our computational graph. 



\section{Experiments}
\label{sec:experiments}

\subsection{Model training}

The network parameters used are listed in Table \ref{tab:parameters}. "Dense" denotes fully connected feedforward layers, and "BN" denotes batch normalization. 
The F0 model was trained with a modified version of CURRENNT \cite{weninger15a-currennt}, available online%
\footnote{\url{https://github.com/TonyWangX/}}.
Excitation models use Keras \cite{chollet2015keras} with Theano \cite{Bastien-Theano-2012} backend 
(code available online\footnote{\url{https://github.com/ljuvela/ResGAN}}). One dimensional convolution layers are used throughout, and paramater $K$ in Conv1D($K$) denotes the number of channels in a layer.

The F0 and pulse regression DNN models were trained with the Adam optimizer \cite{Kingma2014-adam} using early stopping, and frozen after training. 
Unfortunately, there is no established procedure for measuring convergence of GANs. After the first few epochs, the generator starts to produce visually plausible results, while the sound quality varies from epoch to epoch. In the end, we trained the GAN models for a total of 20 epochs, and chose the best model from the last five epochs by informal listening.

\begin{table}[htb]

\caption{Network parameters.}
\label{tab:parameters}
\small
\center
\begin{tabular}{ l }
  \hline
  \textbf{F0 model}, \textit{input}: MFCC(20)   \\
  Dense (256) $\times$ 2, tanh  \\
  BLSTM (128), tanh  \\
  LSTM (128), tanh, feedback link from output \\
  SoftMax(256), 256 class quantization for F0 and voicing  \\
  \textit{output}: F0(1), VUV(1) \\
  \hline
  \textbf{Pulse model}, \textit{input}: MFCC(20), LF0(1), VUV(1) \\
  GRU (50), ReLU, BN, context len = 40 \\
  Dense (400), ReLU, BN \\
  Conv1D (100) $\times$ 4, LReLU, BN, width = 15  \\
  Conv1D (1), LReLU, BN, width = 15 \\
  \textit{output}: Pulse(400)\\
  \hline
  \textbf{GAN generator},  \textit{input}: Noise(400), Pulse(400) \\
  Conv1D (100+1) $\times$ 3, LReLU, BN, width = 15 \\
  Conv1D (1), tanh, BN, width = 15 \\
  \textit{output}: Pulse(400), FFT-of-Pulse(400) \\
  \hline
  \textbf{GAN discriminator}, \textit{input}: Pulse(400), FFT-of-Pulse(400) \\
  Conv1D (64), LReLU, BN, width=7, stride=3 \\
  Conv1D (128), LReLU, BN, width=7, stride=3 \\
  Conv1D (256), LReLU, BN, width=7, stride=3, \textit{peek output} \\
  Conv1D (128), LReLU, BN, width=5, stride=2 \\
  Conv1D (1), LReLU, BN, width=3, stride=2 \\
  \textit{output}: Real/Fake classification (1) \\
  \hline
\end{tabular}

\end{table}

\subsection{Speech material}
We trained two speaker-specific systems using existing SPSS training data. Both speakers are professional UK English voice talents, with "Nick" (male) dataset  comprising 2542 utterances, totaling 1.8 hours, and "Jenny" (female) dataset comprising 4080 utterances, totaling approx.~4 hours. 
A randomly selected set of 100 utterances was kept for testing for both speakers, and the rest were used for training. 16\,kHz sample rate was used throughout the study.

\subsection{F0 objective measures}
\begin{figure} [htb]
  {
  \fontsize{25}{30}
  \resizebox{0.9\linewidth}{!}{\input{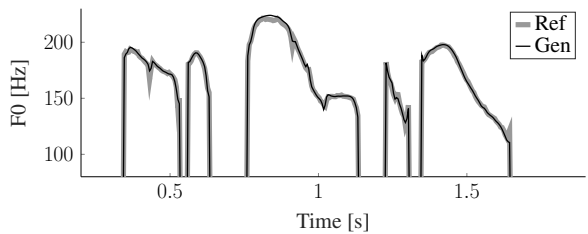}}
  }
\caption{Example of reference and generated F0 contours. }
\label{fig:f0_contour}
\end{figure}

F0 model performance is measured by root-mean-squared error (RMSE) of voiced F0, voicing decision error percentage (VUV error), and correlation coefficient between reference and generated F0 values. Table \ref{tab:objective_measures} lists test set objective measures. An example of a generated F0 contour from "Jenny" test set is shown in Fig.~\ref{fig:f0_contour}.


\begin{table} [htb]
\caption{Objective measures on F0.}
\label{tab:objective_measures}
\small
\center
\begin{tabular}{l | l l l  }
\hline
       & RMSE & VUV error & corr. \\
\hline
 Jenny & 9.3579 & 1.26\,\% & 0.9939 \\
 Nick & 4.2332 & 2.31\,\% & 0.9969 \\
\hline
\end{tabular}
\end{table}

\subsection{Listening test}
\label{sec:listening_test}

Three systems were compared in a DMOS \cite{Itu1996} listening test. All systems use the all-pole envelopes reconstructed from MFCCs, and F0 and voicing information generated by the F0 model. White noise was used for unvoiced excitation in all systems. 
System "Impulse" uses a simple impulse train for voiced excitation. System "DNN" uses the smooth excitation pulses generated by the DNN excitation model, and "GAN" additionally uses the residual GAN noise model. 

\begin{figure} [htb]
\begin{minipage}{.48\linewidth}
  \begin{center}
  \noindent Nick
  \end{center} %
  \vspace{-20pt}
  {
  \Large
  \center
  \resizebox{0.99\linewidth}{!}{
\begin{tikzpicture}

\definecolor{color1}{rgb}{0.172549019607843,0.627450980392157,0.172549019607843}
\definecolor{color0}{rgb}{1,0.498039215686275,0.0549019607843137}

\begin{axis}[
xmin=0.5, xmax=3.5,
ymin=0.8, ymax=5.2,
xtick={1,2,3},
xticklabels={Impulse,DNN,GAN},
tick align=outside,
tick pos=left,
x grid style={lightgray!92.026143790849673!black},
y grid style={lightgray!92.026143790849673!black}
]
\addplot [black, forget plot]
table {%
0.85 3
1.15 3
1.15 5
0.85 5
0.85 3
};
\addplot [black, forget plot]
table {%
1 3
1 1
};
\addplot [black, forget plot]
table {%
1 5
1 5
};
\addplot [black, forget plot]
table {%
0.925 1
1.075 1
};
\addplot [black, forget plot]
table {%
0.925 5
1.075 5
};
\addplot [black, forget plot]
table {%
1.85 4
2.15 4
2.15 5
1.85 5
1.85 4
};
\addplot [black, forget plot]
table {%
2 4
2 3
};
\addplot [black, forget plot]
table {%
2 5
2 5
};
\addplot [black, forget plot]
table {%
1.925 3
2.075 3
};
\addplot [black, forget plot]
table {%
1.925 5
2.075 5
};
\addplot [black, mark=*, mark size=3, mark options={solid,fill opacity=0}, only marks, forget plot]
table {%
2 1
2 1
2 2
2 2
2 2
2 2
2 1
2 1
2 1
2 1
2 1
2 1
2 1
2 1
2 1
2 2
2 1
2 2
2 1
2 2
2 2
2 2
2 1
2 2
2 1
2 2
2 2
2 1
2 1
2 1
2 2
2 2
2 2
2 2
2 1
2 1
2 1
2 1
2 1
2 1
2 1
2 1
2 1
2 2
2 1
2 1
2 2
2 1
2 1
2 2
2 2
2 1
2 1
2 1
2 1
2 1
2 1
2 1
2 2
2 1
2 2
2 2
2 2
2 1
2 1
2 2
2 1
2 2
2 2
2 1
2 1
};
\addplot [black, forget plot]
table {%
2.85 4
3.15 4
3.15 5
2.85 5
2.85 4
};
\addplot [black, forget plot]
table {%
3 4
3 3
};
\addplot [black, forget plot]
table {%
3 5
3 5
};
\addplot [black, forget plot]
table {%
2.925 3
3.075 3
};
\addplot [black, forget plot]
table {%
2.925 5
3.075 5
};
\addplot [black, mark=*, mark size=3, mark options={solid,fill opacity=0}, only marks, forget plot]
table {%
3 1
3 1
3 2
3 2
3 1
3 2
3 1
3 2
3 1
3 1
3 1
3 1
3 2
3 2
3 1
3 2
3 1
3 1
3 2
3 1
3 1
3 2
3 1
3 1
3 1
3 2
3 2
3 2
3 1
3 1
3 1
3 1
3 1
3 1
3 2
3 1
3 2
3 1
3 2
3 1
3 1
3 1
3 2
3 1
3 2
3 1
3 1
3 1
3 2
3 2
3 2
3 1
3 2
3 2
};
\addplot [color0, forget plot]
table {%
0.85 4
1.15 4
};
\addplot [color1, dashed, mark=triangle*, mark size=3, mark options={solid}, forget plot]
table {%
1 3.76666666666667
};
\addplot [color0, forget plot]
table {%
1.85 5
2.15 5
};
\addplot [color1, dashed, mark=triangle*, mark size=3, mark options={solid}, forget plot]
table {%
2 4.29333333333333
};
\addplot [color0, forget plot]
table {%
2.85 5
3.15 5
};
\addplot [color1, dashed, mark=triangle*, mark size=3, mark options={solid}, forget plot]
table {%
3 4.46133333333333
};
\end{axis}

\end{tikzpicture}}
  }
\end{minipage}
\hfill
\begin{minipage}{.48\linewidth}
  \begin{center}
  \noindent Jenny
  \end{center} %
  \vspace{-20pt}
  {
  \Large
  \center
  \resizebox{0.99\linewidth}{!}{
\begin{tikzpicture}

\definecolor{color1}{rgb}{0.172549019607843,0.627450980392157,0.172549019607843}
\definecolor{color0}{rgb}{1,0.498039215686275,0.0549019607843137}

\begin{axis}[
xmin=0.5, xmax=3.5,
ymin=0.8, ymax=5.2,
xtick={1,2,3},
xticklabels={Impulse,DNN,GAN},
tick align=outside,
tick pos=left,
x grid style={lightgray!92.026143790849673!black},
y grid style={lightgray!92.026143790849673!black}
]
\addplot [black, forget plot]
table {%
0.85 2
1.15 2
1.15 5
0.85 5
0.85 2
};
\addplot [black, forget plot]
table {%
1 2
1 1
};
\addplot [black, forget plot]
table {%
1 5
1 5
};
\addplot [black, forget plot]
table {%
0.925 1
1.075 1
};
\addplot [black, forget plot]
table {%
0.925 5
1.075 5
};
\addplot [black, forget plot]
table {%
1.85 3
2.15 3
2.15 5
1.85 5
1.85 3
};
\addplot [black, forget plot]
table {%
2 3
2 1
};
\addplot [black, forget plot]
table {%
2 5
2 5
};
\addplot [black, forget plot]
table {%
1.925 1
2.075 1
};
\addplot [black, forget plot]
table {%
1.925 5
2.075 5
};
\addplot [black, forget plot]
table {%
2.85 4
3.15 4
3.15 5
2.85 5
2.85 4
};
\addplot [black, forget plot]
table {%
3 4
3 3
};
\addplot [black, forget plot]
table {%
3 5
3 5
};
\addplot [black, forget plot]
table {%
2.925 3
3.075 3
};
\addplot [black, forget plot]
table {%
2.925 5
3.075 5
};
\addplot [black, mark=*, mark size=3, mark options={solid,fill opacity=0}, only marks, forget plot]
table {%
3 1
3 1
3 2
3 1
3 2
3 2
3 2
3 2
3 1
3 1
3 1
3 2
3 1
3 2
3 2
3 2
3 1
3 1
3 1
3 2
3 2
3 2
3 2
3 2
3 1
3 1
3 2
3 2
3 2
3 2
3 2
3 2
3 1
3 2
3 2
3 1
3 2
3 2
3 1
3 1
3 2
3 2
3 2
3 2
3 1
3 2
3 1
3 1
3 2
3 2
3 1
3 1
3 1
3 1
3 2
3 1
3 1
3 1
3 2
3 1
3 2
3 2
3 1
3 1
3 2
3 1
3 1
3 2
3 1
3 2
3 2
3 1
3 2
3 2
3 2
3 2
3 2
3 2
3 2
3 1
3 1
3 2
3 2
3 2
3 2
3 1
3 2
3 2
3 1
3 2
3 2
3 1
3 2
3 2
};
\addplot [color0, forget plot]
table {%
0.85 4
1.15 4
};
\addplot [color1, dashed, mark=triangle*, mark size=3, mark options={solid}, forget plot]
table {%
1 3.34266666666667
};
\addplot [color0, forget plot]
table {%
1.85 4
2.15 4
};
\addplot [color1, dashed, mark=triangle*, mark size=3, mark options={solid}, forget plot]
table {%
2 3.92666666666667
};
\addplot [color0, forget plot]
table {%
2.85 5
3.15 5
};
\addplot [color1, dashed, mark=triangle*, mark size=3, mark options={solid}, forget plot]
table {%
3 4.13066666666667
};
\end{axis}

\end{tikzpicture}}
  }
\end{minipage}
\vspace{-10pt}
\caption{DMOS listening test results.}
\label{fig:dmos_results}
\end{figure}
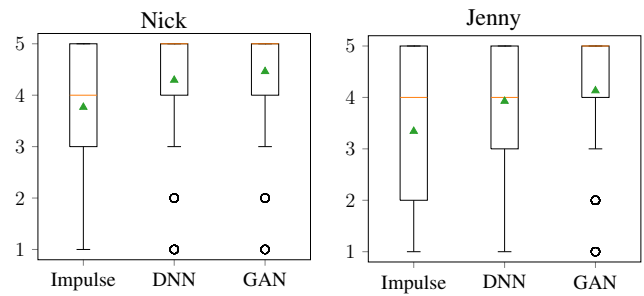

Natural speech signal was used as the reference and the listeners were asked to rate the degradation of the synthetic test sample from 1 (very annoying) to 5 (inaudible).  
The  test was conducted on the CrowdFlower crowd sourcing platform \cite{crowdflower}, where it was made available in English speaking countries, and the top four countries in EF English Proficiency Index ranking \cite{efi-english-proficiency-index}.
Each test case was evaluated by 50 listeners on 15 test set utterances.
Evaluation scores are shown in Fig.~\ref{fig:dmos_results}. Mode value is marked with a horizontal line and the mean value with a triangle. Box and whiskers show 25 and 75 percentile boundaries, respectively.
A Mann--Whitney U-test, with correction for listener and utterance bias  \cite{Rosenberg2017-bias-significance-mos}, found all differences between systems statistically significant. 
Audio samples are available online at {\url{http://tts.org.aalto.fi/mfcc\_synthesis/}}.

\section{Conclusion}
\label{sec:conclusion}

This paper presented a method for speech reconstruction from MFCCs. F0 contours can be generated from MFCCs with high accuracy, using an autoregressive RNN operating on quantized F0 values. The spectral envelope information in MFCCs was recovered by least-squares inversion of the MFCC computation, and a DNN excitation model was trained for the MFCC-derived filters. Additionally, we proposed a residual GAN noise model that can be used to generate a realistic stochastic signal component without explicit parametrization of aperiodicity or similar features.

The listening tests show that a reasonable quality speech reconstruction is obtained already from the MFCC-derived envelope and impulse train excitation with the generated F0. Further improvements were gained with the proposed DNN exitation model and GAN noise model, resulting in high quality speech synthesis from the MFCCs.  

\section{Acknowledgements}
This work was supported by the Academy of Finland (proj.~no.~284671 and 312490.) and MEXT KAKENHI Grant Numbers (26280066, 15H01686, 15K12071, 16H06302).
We acknowledge the computational resources provided by the Aalto Science-IT project.

\newpage
\bibliographystyle{IEEEbib}
\bibliography{refs}

\begin{thebibliography}{10}

\bibitem{davis1980mfcc}
Steven Davis and Paul Mermelstein,
\newblock ``Comparison of parametric representations for monosyllabic word
  recognition in continuously spoken sentences,''
\newblock {\em IEEE Transactions on Acoustics, Speech, and Signal Processing},
  vol. 28, no. 4, pp. 357--366, 1980.

\bibitem{rabiner1993fundamentals-asr}
Lawrence~R Rabiner and Biing-Hwang Juang,
\newblock {\em Fundamentals of {S}peech {R}ecognition},
\newblock PTR Prentice Hall, 1993.

\bibitem{kinnunen2010overview}
Tomi Kinnunen and Haizhou Li,
\newblock ``An overview of text-independent speaker recognition: From features
  to supervectors,''
\newblock {\em Speech Communication}, vol. 52, no. 1, pp. 12--40, 2010.

\bibitem{hansen2015speaker-recognition-tutorial}
John~HL Hansen and Taufiq Hasan,
\newblock ``Speaker recognition by machines and humans: A tutorial review,''
\newblock {\em IEEE Signal Processing Magazine}, vol. 32, no. 6, pp. 74--99,
  2015.

\bibitem{hermansky2013perceptual-properties-asr}
Hynek Hermansky, Jordan~R Cohen, and Richard~M Stern,
\newblock ``Perceptual properties of current speech recognition technology,''
\newblock {\em Proceedings of the IEEE}, vol. 101, no. 9, pp. 1968--1985, 2013.

\bibitem{tokuda1994mel-generalized-cepstrum}
Keiichi Tokuda, Takao Kobayashi, Takashi Masuko, and Satoshi Imai,
\newblock ``Mel-generalized cepstral analysis -- a unified approach to speech
  spectral estimation.,''
\newblock in {\em Proc. ICSLP}, 1994, pp. 18--22.

\bibitem{kinnunen2017non-parallel-vc-ivector}
Tomi Kinnunen, Lauri Juvela, Paavo Alku, and Junichi Yamagishi,
\newblock ``Non-parallel voice conversion using i-vector {PLDA}: Towards
  unifying speaker verification and transformation,''
\newblock in {\em Proc. ICASSP}, 2017, pp. 5535--5539.

\bibitem{boucheron2012low-bitrate-mfcc-codec}
Laura~E Boucheron, Phillip~L De~Leon, and Steven Sandoval,
\newblock ``Low bit-rate speech coding through quantization of mel-frequency
  cepstral coefficients,''
\newblock {\em IEEE Transactions on Audio, Speech, and Language Processing},
  vol. 20, no. 2, pp. 610--619, 2012.

\bibitem{Shao2005-predicting-f0}
Xu~Shao and Ben Milner,
\newblock ``Predicting fundamental frequency from mel-frequency cepstral
  coefficients to enable speech reconstruction,''
\newblock {\em The Journal of the Acoustical Society of America}, vol. 118, no.
  2, pp. 1134--1143, 2005.

\bibitem{milner2007prediction-f0}
Ben Milner and Xu~Shao,
\newblock ``Prediction of fundamental frequency and voicing from mel-frequency
  cepstral coefficients for unconstrained speech reconstruction,''
\newblock {\em IEEE Transactions on Audio, Speech, and Language Processing},
  vol. 15, no. 1, pp. 24--33, 2007.

\bibitem{juvela2016a-high-pitched-excitation}
Lauri Juvela, Bajibabu Bollepalli, Manu Airaksinen, and Paavo Alku,
\newblock ``High-pitched excitation generation for glottal vocoding in
  statistical parametric speech synthesis using a deep neural network,''
\newblock in {\em Proc. ICASSP}, March 2016, pp. 5120--5124.

\bibitem{Bollepalli2017-gan-glottal-excitation}
Bajibabu Bollepalli, Lauri Juvela, and Paavo Alku,
\newblock ``Generative adversarial network-based glottal waveform model for
  statistical parametric speech synthesis,''
\newblock in {\em Proc. Interspeech}, 2017, pp. 3394--3398.

\bibitem{Kaneko2017-gan-postfilter}
Takuhiro Kaneko, Hirokazu Kameoka, Nobukatsu Hojo, Yusuke Ijima, Kaoru
  Hiramatsu, and Kunio Kashino,
\newblock ``Generative adversarial network-based postfilter for statistical
  parametric speech synthesis,''
\newblock in {\em Proc. ICASSP}, March 2017, pp. 4910--4914.

\bibitem{Kaneko2017-learned-similarity}
Takuhiro Kaneko, Hirokazu Kameoka, Kaoru Hiramatsu, and Kunio Kashino,
\newblock ``Sequence-to-sequence voice conversion with similarity metric
  learned using generative adversarial networks,''
\newblock in {\em Proc. Interspeech}, 2017, pp. 1283--1287.

\bibitem{Wang2017-quant-f0}
Xin Wang, Shinji Takaki, and Junichi Yamagishi,
\newblock ``An {RNN}-based quantized {F0} model with multi-tier feedback links
  for text-to-speech synthesis,''
\newblock in {\em Proc. Interspeech}, 2017, pp. 1059--1063.

\bibitem{moulines1995-psola}
Eric Moulines and Jean Laroche,
\newblock ``Non-parametric techniques for pitch-scale and time-scale
  modification of speech,''
\newblock {\em Speech Communication}, vol. 16, no. 2, pp. 175--205, 1995.

\bibitem{young2002htk-book}
Steve Young, Gunnar Evermann, Mark Gales, Thomas Hain, Dan Kershaw, Xunying
  Liu, Gareth Moore, Julian Odell, Dave Ollason, Dan Povey, et~al.,
\newblock {\em The HTK book}, vol.~3,
\newblock Cambridge University, 2002.

\bibitem{mcfee2015librosa}
Brian McFee, Colin Raffel, Dawen Liang, Daniel~PW Ellis, Matt McVicar, Eric
  Battenberg, and Oriol Nieto,
\newblock ``Librosa: Audio and music signal analysis in python,''
\newblock in {\em Proc. of the 14th {P}ython in {S}cience {C}onference}, 2015,
  pp. 18--25.

\bibitem{boucheron2008inversion-mfcc}
Laura~E Boucheron and Phillip~L De~Leon,
\newblock ``On the inversion of mel-frequency cepstral coefficients for speech
  enhancement applications,''
\newblock in {\em Proc. ICSES}. IEEE, 2008, pp. 485--488.

\bibitem{Makhoul1975}
John Makhoul,
\newblock ``Linear prediction: A tutorial review,''
\newblock {\em Proceedings of the IEEE}, vol. 63, no. 4, pp. 561--580, Apr
  1975.

\bibitem{Airaksinen2014}
Manu Airaksinen, Tuomo Raitio, Brad Story, and Paavo Alku,
\newblock ``Quasi closed phase glottal inverse filtering analysis with weighted
  linear prediction,''
\newblock {\em IEEE/ACM Transactions on Audio, Speech, and Language
  Processing}, vol. 22, no. 3, pp. 596--607, March 2014.

\bibitem{talkin2015-reaper}
David Talkin,
\newblock ``{REAPER: Robust Epoch And Pitch EstimatoR},''
  https://github.com/google/REAPER, 2015.

\bibitem{juvela2016b-using-text-and-acoustic}
Lauri Juvela, Xin Wang, Shinji Takaki, Manu Airaksinen, Junichi Yamagishi, and
  Paavo Alku,
\newblock ``Using text and acoustic features in predicting glottal excitation
  waveforms for parametric speech synthesis with recurrent neural networks,''
\newblock in {\em "Proc. Interspeech"}, Sep. 2016, pp. 2283--2287.

\bibitem{Mao2017lsgan}
Xudong Mao, Qing Li, Haoran Xie, Raymond~Y.K. Lau, and Zhen Wang,
\newblock ``Least squares generative adversarial networks,''
\newblock {\em arXiv preprint arXiv:1611.04076v2}, 2017.

\bibitem{weninger15a-currennt}
Felix Weninger,
\newblock ``Introducing {CURRENNT}: The {M}unich open-source {CUDA} recurrent
  neural network toolkit,''
\newblock {\em Journal of Machine Learning Research}, vol. 16, pp. 547--551,
  2015.

\bibitem{chollet2015keras}
Fran\c{c}ois Chollet et~al.,
\newblock ``Keras,'' \url{https://github.com/fchollet/keras}, 2015.

\bibitem{Bastien-Theano-2012}
Fr{\'{e}}d{\'{e}}ric Bastien, Pascal Lamblin, Razvan Pascanu, James Bergstra,
  Ian~J. Goodfellow, Arnaud Bergeron, Nicolas Bouchard, and Yoshua Bengio,
\newblock ``Theano: new features and speed improvements,'' Deep Learning and
  Unsupervised Feature Learning NIPS 2012 Workshop, 2012.

\bibitem{Kingma2014-adam}
Diederik~P. Kingma and Jimmy Ba,
\newblock ``Adam: {A} method for stochastic optimization,''
\newblock in {\em Proc. ICLR}, 2015.

\bibitem{Itu1996}
``{Methods for Subjective Determination of Transmission Quality},''
\newblock Recommendation P.800, ITU-T SG12, Geneva, Switzerland, Aug. 1996.

\bibitem{crowdflower}
{CrowdFlower Inc.},
\newblock ``Crowd-sourcing platform,'' https://www.crowdflower.com/,
\newblock Accessed: 2017-10-25.

\bibitem{efi-english-proficiency-index}
``{EF} {E}nglish proficiency index,'' http://www.ef.com/epi/,
\newblock Accessed: 2017-10-24.

\bibitem{Rosenberg2017-bias-significance-mos}
Andrew Rosenberg and Bhuvana Ramabhadran,
\newblock ``Bias and statistical significance in evaluating speech synthesis
  with mean opinion scores,''
\newblock in {\em Proc. Interspeech}, 2017, pp. 3976--3980.

\end{thebibliography}

\end{document}